\documentclass[nofootinbib, twocolumn, longbibliography,superscriptaddress]{revtex4-2}
\usepackage{orcidlink}
\usepackage{amsmath}
\usepackage{amsthm}
\usepackage{amssymb}
\usepackage{graphicx}
\usepackage{hyperref}
\usepackage{color}
\usepackage{array}
\usepackage[makeroom]{cancel}
\usepackage[capitalize]{cleveref}
\usepackage{changes}
\usepackage{qcircuit}
\usepackage{braket}
\usepackage{bbm}
\usepackage{bm}
\usepackage{booktabs}
\usepackage{titlesec}
\usepackage{comment}
\usepackage{subfigure}
\usepackage{adjustbox}
\DeclareMathOperator{\id}{\mathbbm{1}}

\newcommand{\beq}{\begin{equation}}
\newcommand{\eeq}{\end{equation}}

\usepackage{multirow}
\usepackage{cellspace} 
\usepackage{float}
\makeatletter
\let\newfloat\newfloat@ltx
\makeatother
\usepackage{algorithm}
\usepackage{algpseudocode}

\usepackage{empheq}
\usepackage[most]{tcolorbox}
\newtcbox{\mymath}[1][]{%
    nobeforeafter, math upper, tcbox raise base,
    enhanced, colframe=blue!20!black,
    colback=blue!18, boxrule=0.8pt,
    #1}

\definecolor{MyDarkBlue}{rgb}{0,0.1,0.75}
\hypersetup{pdfborder={0 0 0},colorlinks,breaklinks=true,
  urlcolor={MyDarkBlue},citecolor={MyDarkBlue},linkcolor={MyDarkBlue}
}

\theoremstyle{plain}

\theoremstyle{definition}

\theoremstyle{remark}

\usepackage{nameref}
\newcommand{\figref}[1]{Fig.~\ref{#1}}

\newcommand{\tr}{{\rm Tr}}

\newcommand{\cptp}{{\rm CPTP}}

\algrenewcommand\algorithmicrequire{\textbf{Input:}}
\algrenewcommand\algorithmicensure{\textbf{Output:}}

\newcommand{\bea}{\begin{eqnarray}}
\newcommand{\eea}{\end{eqnarray}}
\newcommand{\be}{\begin{equation}}
\newcommand{\ee}{\end{equation}}
\newcommand{\ba}{\begin{equation}\begin{aligned}}
\newcommand{\ea}{\end{aligned}\end{equation}}
\newcommand{\eqs}[1]{\begin{align}#1\end{align}}

\newtheorem{theorem}{\textbf{Theorem}}
\newtheorem{lemma}{\textbf{Lemma}}

\def\1{\mathds{1}}

\def\id{\mathsf{id}}
\def\mB{\mathcal{B}}

\def\mE{\mathcal{E}}

\def\mG{\mathcal{G}}

\def\mI{\mathcal{I}}

\def\mN{\mathcal{N}}
\def\mO{\mathcal{O}}

\def\mU{\mathcal{U}}

\def\zerostate{|0\rangle\langle 0|}

\def\({\left(}
\def\){\right)}
\def\[{\left[}
\def\]{\right]}

\renewcommand{\theparagraph}{\arabic{paragraph}.}

\titleformat{\paragraph}[runin]
  {\normalfont\itshape}
  {\theparagraph}
  {0.4em}
  {}
  
 \titlespacing{\paragraph}
  {10pt}{0.2\baselineskip}{0.3\baselineskip}
  
\begin{document}
\title{Constant Runtime Error Mitigation via Restricted Evolution}

\author{Gaurav Saxena}
\email{gaurav.saxena@lge.com}
\affiliation{LG Electronics Toronto AI Lab, Toronto, Ontario M5V 1M3, Canada}

\author{Thi Ha Kyaw}
\email{thiha.kyaw@lge.com}
\affiliation{LG Electronics Toronto AI Lab, Toronto, Ontario M5V 1M3, Canada}

\begin{abstract}
\noindent\textbf{\large Abstract}\\
Error mitigation techniques, while instrumental in extending the capabilities of near-term quantum computers, often suffer from exponential resource scaling with noise levels.
To address this limitation, we introduce a novel approach, namely, constant runtime error mitigation by restricted evolution (EMRE). 
Through numerical simulations, we demonstrate that EMRE surpasses the performance of Probabilistic Error Cancellation (PEC) while maintaining constant sampling overhead.
The constant sampling overhead comes at the cost of a small non-zero bias.
We provide a methodology to compute the optimal bias by connecting it to a resource-theoretic measure.
We also evaluate bounds on the bias under different noise models and give exact results for the case of depolarizing and dephasing noise.
Using these exact results, we derive an even more efficient strategy to implement EMRE.
Additionally, we introduce Hybrid EMREs (HEMREs), a continuous family of error mitigation protocols that encompass PEC and EMRE as special cases. HEMREs offer a tunable bias parameter, enabling a trade-off between sample complexity and error reduction. 
The numerical evidence suggests the scalability and practicality of our proposal.
Hence, our error mitigation protocols provide flexibility in balancing error mitigation with computational overhead, catering to practical application requirements of near-term and early-fault tolerant quantum devices.
\end{abstract}
\maketitle

\noindent\textbf{\large Introduction}\\
\noindent Practical quantum computational advantage is one of the holy grails that the entire quantum information community aims to achieve.
It has recently been pointed out that material science and chemistry are two potential areas of applications where practical quantum advantage might be unleashed \cite{Hoefler2023Apr}.
It is believed that such an event will occur at the level of early fault-tolerant quantum computing that involves a substantial number of qubits and quantum gates \cite{Lee2021Jul} that the current near-term quantum hardware does not possess.
The design of such powerful quantum computers will require sustained effort in the field for a very long time, making the possibility of a ``quantum winter" a real concern.
In order to sustain the momentum of top-notch research and continual funding support, the quantum information community must show some form of practical quantum advantage \cite{Daley2022Jul,Li2023Mar} using near-term or pre-fault-tolerant quantum hardware.

On the other hand, the near-term quantum hardware is prone to undesired noise. 
Unlike fault-tolerant quantum algorithms, most near-term quantum algorithms \cite{Bharti_2022,Tilly2022Nov,Cerezo2021Sep} are short-depth and come in the form of a hybrid quantum-classical feedback loop.
Since quantum circuits do not take up a heavy load in the entire calculation, a higher noise budget is allowed in the hybrid algorithms.
A common feature of such hybrid approaches is the variational method \cite{Farhi2014Nov,Peruzzo2014Jul,Wecker2015Oct,McClean2016Feb,Kyaw2023Oct}. 

Typically, characterizing target quantum systems often involves constructing trial wavefunctions with a substantial yet manageable number of variational parameters. These parameters are then optimized to minimize the energy of the system by invoking a classical computer in the form of a feedback loop. Implicitly, this approach leverages the user's intuition and knowledge of the target system to select a parameter space that, while extensive, is significantly smaller than the full Hilbert space. The latter, of course, grows exponentially with the number of particles in the system.
One major computational routine there is to compute the expectation value of some physical observables \cite{Bravyi2021Mar}.
Even with such hybrid quantum-classical algorithms, current noisy quantum hardware could produce unreliable measurement outcomes of the physical observables.
What we recently showed in Ref.~\cite{Saxena2024May} is that due to the noisy nature of quantum hardware, in order to see some meaningful results in quantum simulations, the quality of the quantum hardware should be below certain gate fidelities, thereby putting a hard limit on the practicality of the noisy devices.

Recently, a handful of quantum error mitigation (QEM) techniques have been proposed \cite{Li2017Jun,Temme2017Nov,Endo2018Jul,Bonet-Monroig2018Dec,Giurgica-Tiron2020Oct,Sun2021Mar,Shaw2021May,Lowe2021Jul,Koczor2021Sep,Wang2021MitigatingQE, Suzuki2022Mar,Bultrini2023Jun, liu2024virtualchannelpurification} to improve the expectation values in noisy quantum hardware.
The hope is that with the aid of QEM, one would be able to carry out near-term quantum algorithms effectively without needing to worry about the experimental noises. 
Intuitively, such an approach would allow us to increase the quantum system size systematically, thereby potentially crossing a threshold where any best classical computing algorithm would not be able to catch up with quantum ones due to either exponentially large Hilbert space or highly entangled many-body quantum ground state structures.
Hence, the term ``quantum utility era" \cite{Kim2023Jun}. 
A caveat though is that most of the existing quantum error mitigation schemes require either exponential samplings or exponential copies of quantum states, with respect to the amount of noise, to be able to attain resolvable and reliable statistics of the measurement outcome \cite{Cai2023Dec}.
In fact, there exist universal performance limits, suggesting that the sampling overhead scales exponentially with the circuit depth given a desired computational accuracy \cite{PhysRevLett.131.210602, Takagi2022Sep,PhysRevLett.131.210601}. Even at relatively shallow circuit depth, a super-polynomial number of samplings is needed \cite{Quek2024Jul}.

Our primary objective is to mitigate the exponential computational overhead associated with quantum error mitigation protocols while preserving reasonable accuracy.
Recognizing the inherent trade-off between computational efficiency and statistical precision, we propose methods that intentionally introduce biases into the expectation value estimates. In contrast to probabilistic error cancellations \cite{Temme2017Nov,Endo2018Jul}, our approach requires significantly less sampling overhead to achieve comparable results while removing the exponential sampling overhead. 
In fact, we formally prove that our proposal can estimate expectation values with constant sampling complexity.
Moreover, unlike zero-noise extrapolation \cite{Li2017Jun,Temme2017Nov}, our techniques avoid extrapolation procedures, thereby eliminating the need for heuristic assumptions.

We develop a constant runtime quantum error mitigation protocol by approximating each quantum gate present in the quantum circuit with an implementable noisy operation, thereby restricting the quantum dynamical evolution of the input quantum state. Hence the name error mitigation by restricted evolution (EMRE).
The main idea of EMRE is inspired by the generalized robustness measure defined for quantum operations in the context of resource theory of channels~\cite{Takagi2021Aug,Liu2020Feb,Liu2019Apr,Saxena2020Jun}.
At the heart of EMRE, the generalized robustness is used to find the closest implementable (noisy) circuit to the ideal quantum circuit.
Using the new implementable circuit, we show that we can estimate the ideal expectation value closely using a constant sampling overhead.
The price we pay is that the estimated expectation value comes with a small non-zero bias.
We also establish an analytical relationship between the bias and the generalized robustness.

Our results have wide-ranging consequences for near-term quantum experiments.
First, we show that, with our proposed error mitigation strategy, there exist exact analytical expressions of generalized robustness for the depolarizing and dephasing noise models.
That means mitigating noise in these cases just involves a simple post-processing in the form of multiplication by a constant factor (which depends on the strength of the noise) followed by a few checks, thereby showing that no extra classical or quantum resource is required for error mitigation.
Second, we uncover a family of error mitigation protocols by combining PEC and EMRE. 
By doing so, we are able to incorporate desirable features from PEC and EMRE. 
We call them the hybrid EMREs or HEMREs.
The key feature of HEMREs is that it takes as an input the maximum tolerable bias from a user to generate an error mitigation technique that uses much less sampling overhead than PEC. 
The resulting bias is less than or equal to the chosen bias. 
We believe such strategy is greatly beneficial to those with limited QPU hours during the quantum cloud execution.\\

\noindent\textbf{\large Results}\\
\textbf{Error Mitigation by Restricted Evolution (EMRE)}\\
\noindent Let an $n$-qubit quantum circuit with depth $D$ be represented by $\mU$ as
$$\mU := \mU_D\circ\cdots\circ\mU_2\circ\mU_1$$
where each $\mU_i$ represents the $i$th layer of the circuit acting on the $n$ qubits.
For each layer, we can find a number $s_i\geq 1$ and an implementable operation $\mB_i$ (or a convex combination thereof) such that $s_i\mB_i\geq \mU_i\,$,
i.e., $s_i J^{\mB_i} - J^{\mU_i}\geq 0$ where $J^{\mB_i},\, J^{\mU_i}$ are the Choi matrices of the operations $\mB_i$ and $\mU_i$, respectively.
The above inequality implies that there exist a set of quantum channels $\{\mN_i\}$ such that for each respective $\mU_i$, it holds that
\eqs{\label{eq:generalized_quasi_decomposition}
\mU_i = s_i\mB_i - (s_i -1)\mN_i\,.
}
The optimal $s_i-1$ can be thought of as generalized robustness, inspired by the generalized robustness measure from dynamical resource theories \cite{Takagi2021Aug,Liu2020Feb,Liu2019Apr,Saxena2020Jun}. (In-depth discussions on the generalized robustness are provided in the methods subsection.)
Let us call the decomposition in Eq.~\eqref{eq:generalized_quasi_decomposition} as the generalized quasi-probability decomposition (generalized QPD or GQPD) where the terms with negative coefficients need not be implementable operations.
Using the above decomposition for all gates in the circuit, we can represent the whole circuit $\mU$ as follows
\eqs{\label{eq:full_circuit_approximation}
\mU = s\mB - (s-1)\mN}
where 
\eqs{s &= \Pi_{i=1}^D s_i\label{eq:s_defn}\,,\\
\mB&= \mB_D\circ\cdots\circ\mB_2\circ\mB_1\,,}
and $\mN$ is the quantum channel that consists of all other combinations.
Using this decomposition, we aim to output an estimate $E$ of the expectation value $\tr[\mO\mU(|0\rangle\langle 0|)]$ of some observable $\mO$ such that
\eqs{\label{eq:actual_bias_eqn}
|E - \tr[\mO\mU(|0\rangle\langle 0|)]|\leq b\,,} 
where $b$ represents a small bias.

\begin{figure*}[h]
        \centering
        \includegraphics[width=1\linewidth]{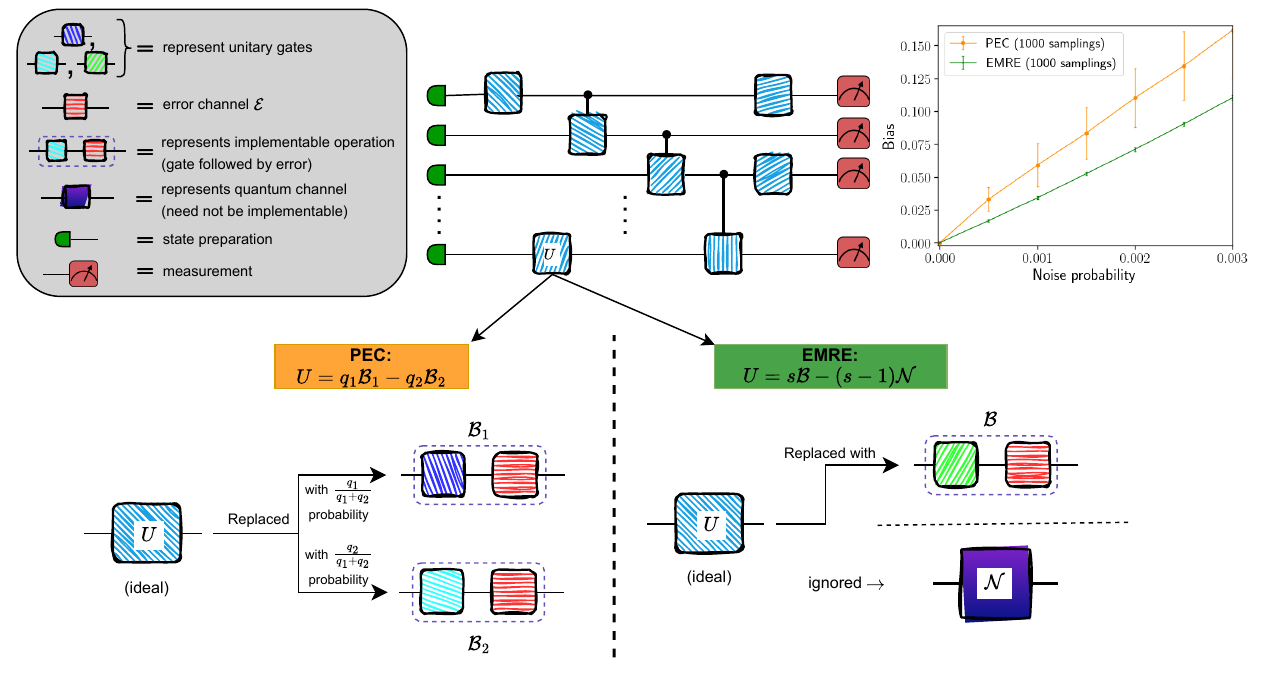}
        \caption{\textbf{Error Mitigation by Restricted Evolution and illustrative comparison with Probabilistic Error Cancellation.} Illustration of mitigating errors by restricting the evolution of the quantum state, outlining the difference between EMRE and PEC in terms of circuit construction and expected outcomes. Performing error mitigation by restricted evolution (or EMRE) requires a constant sample size as opposed to PEC which has an exponential sampling overhead. In practical applications, PEC is deployed by reducing the sampling budget. Under such limited‑sampling conditions, we numerically demonstrate that EMRE outperforms PEC, as shown in the inset, where both methods use 1000 samples. We would like to point out the stability of our proposed EMRE as apparent from very small error bars across different noise probabilities. The inset comparison is intended for illustrative purposes and conveys the following key observations: under a \textit{finite sampling constraint}, which is common in practice, PEC exhibits a larger mean bias than EMRE at low noise levels; EMRE maintains a significantly smaller variance, and consequently, EMRE provides more reliable estimates even at limited sample sizes.
        We note that the present figure displays both the bias and the standard deviation to compactly convey more information; in practice, however, the standard deviation is relevant only in conjunction with the corresponding expectation value.}
        \label{fig:emre_key_flowchart}
\end{figure*}

\begin{figure}[h]
    \centering
    \includegraphics[width=0.75\linewidth]{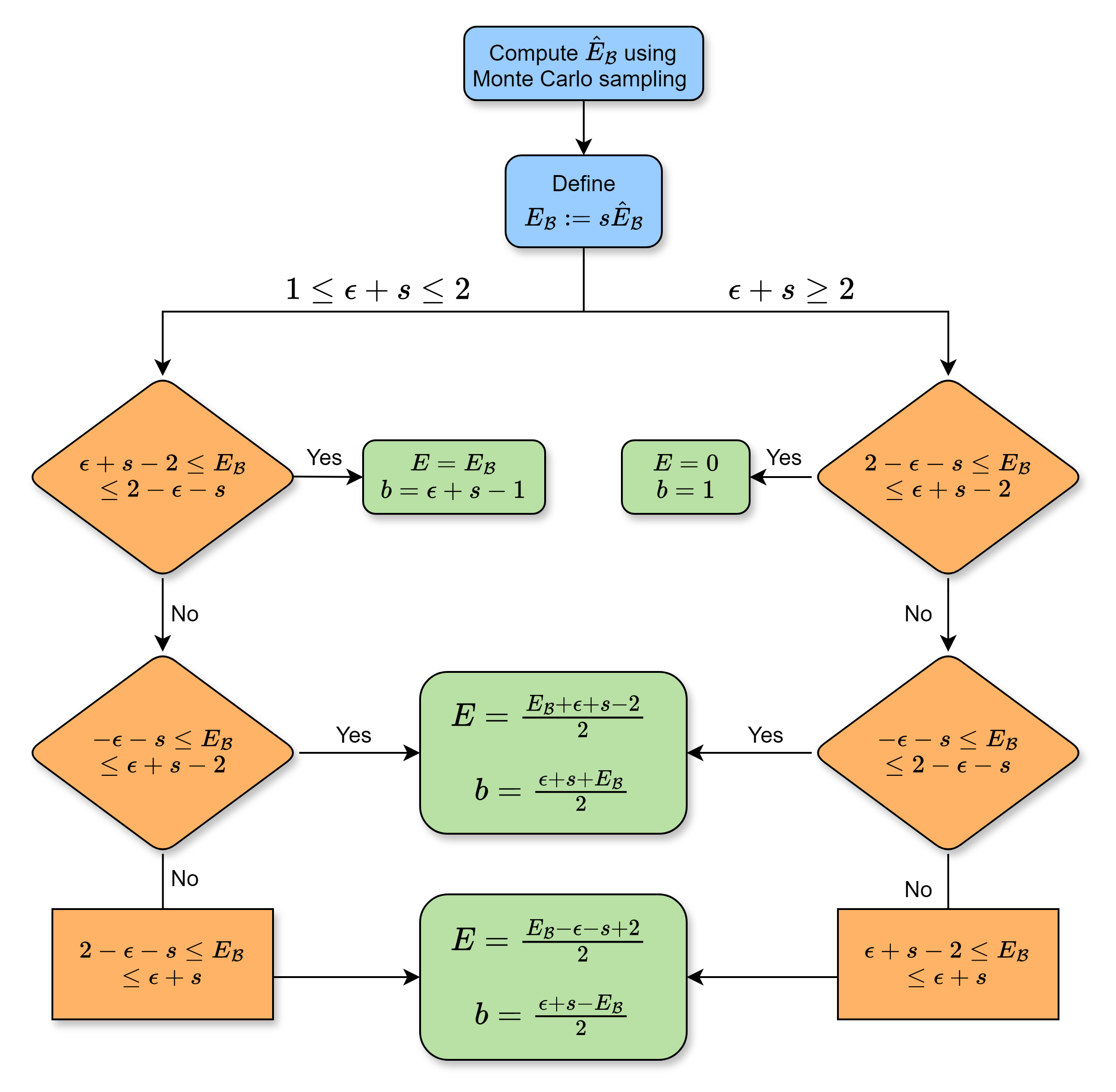}
    \caption{\textbf{EMRE post-processing flowchart.} Flowchart to get the final expectation value from EMRE.}
    \label{fig:emre_flowchart_for_expvals}
\end{figure}

To reduce the sampling overhead to estimate the expectation value $E$, we approximate each unitary $\mU_i$ with $s_i\mB_i$. Practically, this means that we replace the $i$-th operation (or the $i$-th layer) in the circuit with $\mB_i$ and multiply the estimate after the measurement with $s_i$ (see the illustration in Fig.~\ref{fig:emre_key_flowchart}).
Therefore, we approximate the whole circuit with implementable operations as follows:
\eqs{\label{eq:approximating_circuit}
\mU \approx s\mB=s\mB_D\circ\cdots\circ\mB_2\circ\mB_1\, .}
Approximating the circuit in this way, we can say that we restrict the evolution of the input state under the ideal circuit $\mU$ to the implementable circuit $\mB$.
The price we pay is that the bias $b$ (in Eq.~\eqref{eq:actual_bias_eqn}) will always be finite, which comes from the approximation of $\mU$ in Eq.~\eqref{eq:approximating_circuit} by restricting the unitary circuit with an implementable (noisy) circuit.
The proposed algorithm can be considered a direct application of the classical simulation algorithm presented in Ref.~\cite{PhysRevA.106.042422}.
Next, we discuss in detail how to estimate the expectation value $E$ and the bias $b$ from the above approximation (also refer to Fig.~\ref{fig:emre_flowchart_for_expvals}).

Let $c\ll1$ be a fixed small positive constant.
Using Monte Carlo samplings of various $\mB_i$'s in the circuit, we compute $\hat{E}_{\mB}$, the estimate of the expectation value $\tr[\mO\mB(\zerostate)]$ such that 
\eqs{|\hat{E}_{\mB}-\tr[\mO\mB(\zerostate)]|\leq c\,.}
Note that sampling of $\mB_i$'s will be efficient, since we are sampling from a probability distribution. This is the advantage that we gain by restricting the decomposition to its positive parts.
Assuming the probability of failure of the Monte Carlo algorithm to be $p_{\rm fail}$, from Hoeffding's inequality we get the sampling overhead $M_{\textrm{EMRE}}$ required to estimate $\tr[\mO\mB(\zerostate)]$ to be
\tcbset{highlight math style={boxsep=1mm,colback=blue!25!red!25!white}}
\begin{empheq}[box=\tcbhighmath]{equation}\label{eq:our_sampling_overhead}
    M_{\textrm{EMRE}} = \frac{2}{c^2}\ln\left(\frac{2}{p_{\rm fail}}\right),
\end{empheq}
which is a constant.
This is the key result of our paper.

From the estimate $\hat{E}_{\mB}$, we define $E_{\mB}$, the estimate of $s\tr[\mO\mB(\zerostate)]$, as $E_{\mB} := s\hat{E}_{\mB}$.
Then, by defining another quantity
\eqs{\label{eq:epsilon}
\epsilon:= cs\,,}
we get the following bound on the estimate of $s\tr[\mO\mB(\zerostate)]$:
\eqs{\label{eq:EmB_approximation}
|E_{\mB} - s\tr[\mO\mB(|0\rangle\langle 0|)]|\leq \epsilon.}
From~\eqref{eq:EmB_approximation}, we know that either of the following two equations hold.
\eqs{0\leq E_{\mB}-s\tr[\mO\mB(|0\rangle\langle 0|)]&\leq \epsilon,\; {\rm or} \\
0\leq s\tr[\mO\mB(|0\rangle\langle 0|)]- E_{\mB}&\leq \epsilon\;.}
By using~\eqref{eq:full_circuit_approximation}, we can write
\eqs{\tr[\mO\mU(\zerostate)]= s\tr[\mO\mB(\zerostate)]- (s-1)\tr[\mO\mN(\zerostate)]\,.}
Therefore, we get the following bounds on the ideal expectation value.
\eqs{\label{eq:bounds_on_ideal_expval}E_{\mB}-\epsilon-(s-1)\leq \tr[\mO\mU(\zerostate)]&\leq E_{\mB} + \epsilon+(s-1) .}
Since we have assumed $\mO$ to be a Pauli observable, we trivially have $|\tr[\mO\mU(\zerostate)]|\leq 1$.
(In case $\mO$ is not a Pauli observable and if we denote the maximum eigenvalue of $\mO$ by $o_{\max}$, then we have $|\tr[\mO\mU(\zerostate)]|\leq o_{\max}$.)
Now using the above fact and the bound presented in Eq.~\eqref{eq:bounds_on_ideal_expval}, we get different expectation value estimates for different cases of the known values, $E_{\mB},\, s,$ and $\epsilon$, which we present below. The details of this analysis for different cases are provided in \textit{Supp. Info. Section 8}.

For the case when $1\leq \epsilon+s\leq 2$ and $\epsilon+s-2 \leq E_{\mB} \leq 2-\epsilon - s$, the estimate of the expectation value, $E$, is given by $E_{\mB}$ and the (maximum) bias $b$ is
\begin{empheq}[box=\mymath]{equation}\label{eq:EMRE_bias_non-trivial_case}
    b = \epsilon+s-1\,.
\end{empheq}
Note that in the above equation, the bias is a linear function of $s$ which in itself is the product of all the $s_i$'s of the individual gates in the circuit. For simplicity, if we assume that all the $s_i$'s are almost equal to each other, it can be seen that $s$ grows exponentially with the number of gates in the circuit, and so does the bias.

Next, if $1\leq \epsilon+s\leq 2$ and $-\epsilon -s\leq E_{\mB} \leq \epsilon+s-2$, or if $\epsilon+s\geq 2$ and $-\epsilon -s\leq E_{\mB} \leq 2-\epsilon-s$, then we have the following relation:
\eqs{\left|\tr[\mO\mU(\zerostate)] -  \frac{E_{\mB}+\epsilon+s-2}{2}\right|\leq \frac{E_{\mB}+\epsilon+s}{2}\,,}
from which we infer the estimate of the expectation value to be $(E_{\mB}+\epsilon+s-2)/2$ and the maximum bias $b$ is $(E_{\mB}+\epsilon+s)/2\,.$

Similarly, when $1\leq \epsilon+s\leq 2$ and $2-\epsilon-s\leq E_{\mB}\leq \epsilon+s$, or when $\epsilon+s\geq 2$ and $\epsilon+s-2\leq E_{\mB}\leq \epsilon+s$, then we get the following relation:
\eqs{\left|\tr[\mO\mU(\zerostate)] -  \frac{E_{\mB}-\epsilon -s+2}{2}\right|\leq \frac{\epsilon+s-E_{\mB}}{2}\,,}
from which we infer the estimate of the expectation value to be $(E_{\mB}-\epsilon-s+2)/2$ and the maximum bias $b$ is $(\epsilon+s-E_{\mB})/2\,.$

Lastly, a trivial case exists when $\epsilon+s \geq 2$ and
$2-\epsilon-s\leq E_{\mB} \leq \epsilon+s-2$.
In such a case, the algorithm will output the estimate $E$ equal to zero and the bias $b$ equal to one.

A complete flowchart to estimate the expectation value from EMRE under various trivial and nontrivial conditions is presented in Fig.~\ref{fig:emre_flowchart_for_expvals}.\\

\noindent\textbf{Hybrid Error Mitigation by Restricted Evolution (HEMRE)}\\
EMRE estimates the expectation value of an observable using only a constant sampling overhead and is very reliable for quantum gates with low noise probability of the current quantum hardware \cite{Gupta2024Jan,Bluvstein2024Feb,Abdurakhimov2024Aug,Acharya2024Aug}.
However, for noisy hardware with higher noise probabilities, bias from the EMRE's expectation value increases.
To tackle this problem along with an intention to suppress the sampling overhead required by PEC, we develop a hybrid approach by combining EMRE and PEC, resulting in the hybrid error mitigation by restricted evolution (HEMRE) with better sampling overhead than the PEC and better bias than the EMRE.

\begin{figure*}[t]
    \centering
    \includegraphics[scale = 0.8]{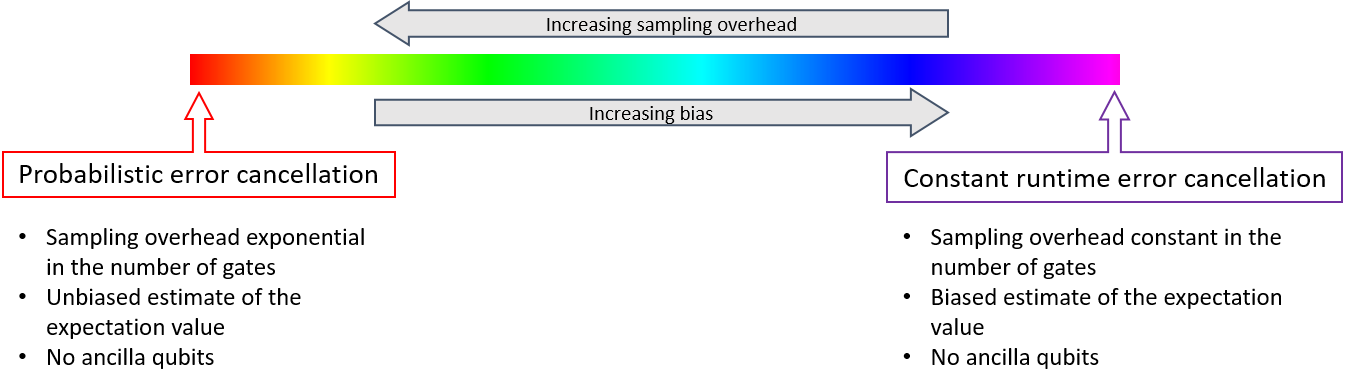}
    \caption{\textbf{Error Mitigation protocols spectrum from PEC to EMRE. } Family of error mitigation protocols color coded in the spectrum where PEC sits at one end and EMRE at the other end. Everything in between is HEMRE.}
    \label{fig:family}
\end{figure*}

There are two ways to construct such a hybrid algorithm.
In the first approach, the strategy is to reduce the sampling overhead given a desired minimum precision in estimating the result.
In this approach, the maximum allowed bias can be treated as a parameter, and we can view EMRE, HEMRE, and PEC together in a spectrum as shown in Fig.~\ref{fig:family} where EMRE and PEC lie at the extreme ends.
In the second approach, the strategy is to minimize the bias given a fixed sampling overhead.
This is achieved by using the Hoeffding's inequality that provides a connection between the sampling overhead and the bias.
Below, we provide the details of the algorithm of the former approach. 
The algorithm for the latter case can be constructed in a similar fashion.
Importantly, for both strategies, one must know in advance both a quasi-probabilistic and a generalized quasi-probabilistic decomposition of each unitary gate $\mU_i$ present in the circuit.
Trivially, the best result is obtained when the decompositions are optimal.

Let us look at HEMRE in full glory and how one can combine EMRE and PEC, given a maximum tolerable bias to reduce the sampling overhead.
The question at hand is which gates in the circuit need to be approximated to the closest implementable gates such that the bias in the result does not exceed $\Delta_{\rm fixed}$, the allowed fixed bias for estimating the expectation value.
In the previous section/discussion on EMRE, we found that the bias depends on the generalized robustness of the gates in the circuit.
We take advantage of this information to answer the above question for HEMRE.
In other words, we need to put a constraint on the product of the generalized robustness of the gates that will be approximated.
Using this constraint, we can find the gates that obey the constraint and we can approximate them to their closest implementable gates or a convex combination thereof.

For the sake of simplicity in the rest of this subsection, let us refer to the factor $s$ (see Eq.~\eqref{eq:s_defn}) as the generalized robustness instead of $s-1$,
and let $\gamma$ denote the product of the robustness of all gates in the circuit.
From Eq.~\eqref{eq:our_sampling_overhead}, Eq.~\eqref{eq:epsilon} and discussions in Ref.~\cite{Temme2017Nov}, the sampling overhead is proportional to the square of the product of the robustness or the generalized robustness of the gates in the circuit.
The key idea here is that we selectively choose to keep the quasi-probabilistic decomposition (QPD) of some gates and for the remaining others, we approximate them with the implementable gates as obtained from their generalized quasi-probability decomposition.
Of course, this selection depends on the input $\Delta_{\rm fixed}$ that a user desires.
By using the Monte-Carlo sampling technique to estimate the expectation value upto $\epsilon~(=cs)$ precision and with success probability $1-p_{\rm fail}$, the sampling overhead $M_{\rm HEMRE}$ is found to be
\eqs{M_{\rm HEMRE} = \Theta\left(\frac{2s_{\rm incl}^2 \gamma^2_{\rm incl}}{\epsilon^2}\ln\left(\frac{2}{p_{\rm fail}}\right)\right)\,,}
where $\gamma_{\rm incl}$ is the total robustness of the gates whose quasi-probability decomposition are chosen for the HEMRE, $s_{\rm incl}$ is the total generalized robustness of the (remaining) gates whose generalized quasi-probability decomposition are chosen,
and $\Theta(\cdot)$ denotes the asymptotic big-Theta notation.
The bias $\Delta$ or the deviation from the actual expectation value can be calculated in a similar way as in EMRE, by considering $s_{\rm incl}$ instead of $s$.
Since in HEMRE we first need to choose the gates that will be approximated, we need to consider only the case where $1\leq \epsilon + s_{\rm incl} \leq 2$ and where the bias is equal to $\epsilon + s -1$ to put a constraint on $s_{\rm incl}$. In all other cases of EMRE, the bias depends on the estimated $E_{\mB}$.
Additionally, we need this bias value to not be more than the user's input, $\Delta_{\rm fixed}$.
Therefore, we get the following constraint on the product of the generalized robustness, $s_{\rm incl}$, of the approximated gates to be
\eqs{\label{eq:bound_on_s_incl}
s_{\rm incl} \leq \Delta_{\rm fixed} +1 - \epsilon\,.}
Using this constraint, we can choose which gates to be approximated.
For the remaining gates, we will use their full quasi-probabilistic decomposition.
There are several ways to realize this strategy. 
We have listed a number of them in the \textit{Supp. Info. Section 5}.
Below, we provide the details of one such method which maximizes the number of gates to be approximated.

\begin{algorithm}
\caption{An algorithm to approximate maximum number of gates for HEMRE}\label{alg:selecting_gates_for_HEMRE}
\begin{algorithmic}[1]
\Require i. Maximum tolerable bias, $\Delta_{\rm fixed}$,\newline
$\:\quad$ii. Gates' information (as in Table~\ref{table:hemre_gates_gen_robustness}) sorted in increasing order of the generalized robustness.
\begin{flushleft}
\textbf{Pre-computation:}
\end{flushleft}
    i. Create a frequency dictionary, say $fr$ = \{gate : frequency\}, containing the frequency of occurrence of corresponding gates in the circuit.\newline
    ii. Create another dictionary, $gr$=\{gate : generalized robustness\}, containing the generalized robustness of corresponding gates.
\Ensure All the gates that need to be replaced, and the corresponding total $s_{\rm incl}$.

\State $s_{\rm incl} \gets 1$
\For{gate \textbf{in} ${\rm unique\_gates}$} 
    \If{$s_{\rm incl}*(gr[{\rm gate}])^{fr[{\rm gate}]}\leq \Delta_{\rm fixed}- \epsilon + 1$}
        \State $s_{\rm incl} \leftarrow s_{\rm incl}*(gr[{\rm gate}])^{fr[{\rm gate}]}$
        \State Approximate all occurrences of `gate' in the circuit
    \Else
        \State $m = \left\lfloor \log\left(\frac{\Delta_{\rm fixed}- \epsilon + 1}{s_{\rm incl}}\right)/\log\left(gr[{\rm gate}]\right) \right\rfloor$
        \State $s_{\rm incl} \leftarrow s_{\rm incl}*(gr[{\rm gate}])^m$
        \State Approximate $m$ occurrences of `gate' in the circuit
        \State Return 
    \EndIf
\EndFor

\end{algorithmic}
\end{algorithm}
To maximize the number of gates to be approximated to the closest (convex combination of) implementable gate(s) for HEMRE, we first need to identify the unique gates in the circuit and compute their generalized robustness.
Using this information, we can construct a table similar to Table~\ref{table:hemre_gates_gen_robustness}.
We can then sort the table according to the generalized robustness such that $s_1\leq s_2 \leq s_3 \leq \cdots$.
Using this sorted table, we can easily find which gates and how many of those gates are required such that the product of their generalized robustness meet the criterion in~\eqref{eq:bound_on_s_incl}.
We also provide an algorithm (see Algorithm~\ref{alg:selecting_gates_for_HEMRE}) to select the maximum number of gates to be approximated such that approximating one more extra gate will violate the constraint of~\eqref{eq:bound_on_s_incl}.
For the remainder of the gates, we consider their quasi-probabilistic decomposition.

The reason for taking this approach of first restricting gates with the smallest generalized robustness over other approaches is that this way we are not losing much information per approximated gate.
This happens because, for the gate with lesser generalized robustness, the ratio between the norm of the positive and the negative part norm will be bigger as compared to that of a gate with larger generalized robustness.
In other words, this implies that for the gate with smaller generalized robustness, the quantum operation that we get by normalizing the positive part of the gate's decomposition is closer to the original gate (in diamond norm) as compared to the quantum operation that we get by normalizing the positive part of the gate with higher generalized robustness.
Thus, when we restrict a gate with a larger generalized robustness (to its positive part), the approximation is not too close to the original gate, and so, it is better to restrict the gates with a smaller generalized robustness.

Lastly, we would like to highlight that with EMRE the number of shots to estimate the expectation value remains a constant irrespective of the circuit size. 
With increasing circuit size and noise probability, the bias does increase, and so to get the best estimate of the ideal expectation value, the optimal generalized QPD must be used.
This also implies that a performance guarantee for large circuit sizes with large noise probabilities is hard to obtain with EMRE.
However, in such cases, we can use HEMRE which takes into account this performance guarantee into its algorithm by taking as input the maximum tolerable bias and producing a result by using a smaller runtime than PEC.\\

\noindent\textbf{Numerical Analysis}\\
To numerically demonstrate our ideas, we use a SWAP-test to test how much the $n$-qubit $|{\rm GHZ}\rangle$ state differs from the $|0\rangle^{\otimes n}$ state. The ideal value of this test for the given states is $0.5$.
$|{\rm GHZ}\rangle$ state preparation is an important experimental benchmark for quantum processors, which serves not only as a test of device-wide entanglement, but also as a crucial resource in many quantum algorithms~\cite{Cruz2019AQT, Graham2022Apr,Bluvstein2024Feb, Omran2019Aug, RevModPhys.89.035002}.

We model the noisy circuit by appending the noisy channel after each gate instance in the circuit.
We consider two practically significant noise models: the partially depolarizing noise and the inhomogeneous Pauli noise model.
In the main text, we present the results of the depolarizing noise with a circuit with 7 qubits and 140 gates. 
Additional results from numerical simulations, including depolarizing noise for larger circuits (demonstrating scalability) and inhomogeneous Pauli noise for the above circuit (illustrating the consistency of results across general noise models), are provided in \textit{Supp. Info. Section 6}.
We then deploy the error mitigation protocols and estimate how much the two states differ from each other. The difference of the obtained value from 0.5 is the bias and we analyze how the bias changes as the noise probability changes for different mitigation protocols.

To implement the SWAP-test, we use the C-SWAP circuit composed of $T$, $T^{\dagger}$, Hadamard, and controlled-NOT gates (see Fig. 4 Sec.~XI of Ref.~\cite{Endo2018Jul} for the circuit). 
The circuit consists of 7 qubits and 140 gates.
We assume some noise to be acting locally after each gate to mimic some forms of experimental conditions.
By simulating such a noisy circuit, we then numerically compute the expectation value of $\langle Z \rangle$ on the ancilla qubit of the C-SWAP circuit which quantifies how much the two input states differ. 
We obtain the estimate of the expectation value for different noise probabilities.
For each noise probability, we estimate the expectation values for five cases -- first: no error mitigation, second: PEC with $1,000$ samplings, third: EMRE with $1,000$ samplings implemented by restricting the QPD used for PEC, fourth: HEMRE with $1,000$ samplings and by specifying the tolerable bias to be the mean bias from the PEC's numerical results at the previous noise probability, and fifth: EMRE performed using the optimal generalized QPD analytically obtained in Theorem~\ref{thm:gen_robustness_under_depolarizing_noise} or Lemma~\ref{lem:gen_rob_under_probabilistic_noise}.
For the last case, where we use the analytical results that the optimal implementable operations are the noisy gates themselves (for the partially depolarizing noise and the probabilistic noise), the first step in the algorithm in~\figref{fig:emre_flowchart_for_expvals} is taken to be the mean of the noisy expectation values.
Due to this, there are no error bars for this optimal plot.
The next steps are the multiplication of this mean (of the noisy results) by $s$ followed by some checks as given in~\figref{fig:emre_flowchart_for_expvals} to output the estimate of the expectation value. See the methods section under the `Bounds on generalized robustness' for more technical details.

\begin{figure*}[h]
    \centering
    \includegraphics[width=0.8\linewidth]{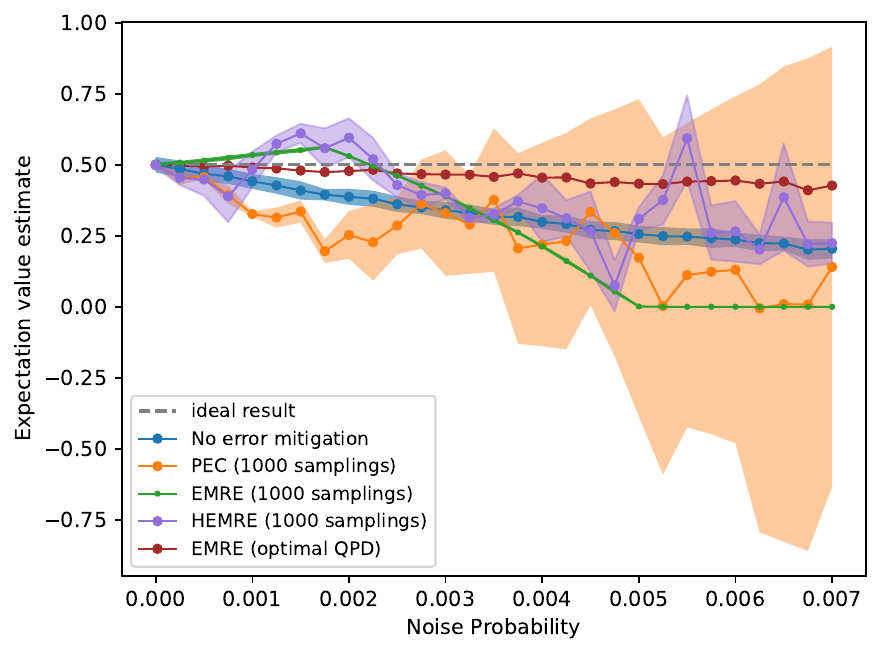}
    \caption{\textbf{Comparison of error mitigation protocols for a noisy C‑SWAP circuit.} Comparison of the expectation value estimates obtained for different error mitigation protocols applied to the C-SWAP circuit under the partially depolarizing noise. EMRE is performed using two methods. First, we execute EMRE by restricting the QPD of each gate in the circuit used for PEC to its positive part (green line). Second, we used the optimal generalized QPD (optimal GQPD) and use the checks provided in~\figref{fig:emre_flowchart_for_expvals} to output the final mitigated values (brown line).
    For HEMRE, we varied the tolerable bias and set it to be equal to the mean bias we get from PEC as the probability of noise increases.
    For PEC, EMRE (green line), and HEMRE, we sampled $1,000$ times from the respective decompositions used in each technique to make a fair comparison in estimating each expectation value.
    }
\label{fig:Comparison_of_expval_under_depol}
\end{figure*}

\begin{figure*}[h]
    \centering
    \includegraphics[width=0.7\linewidth]{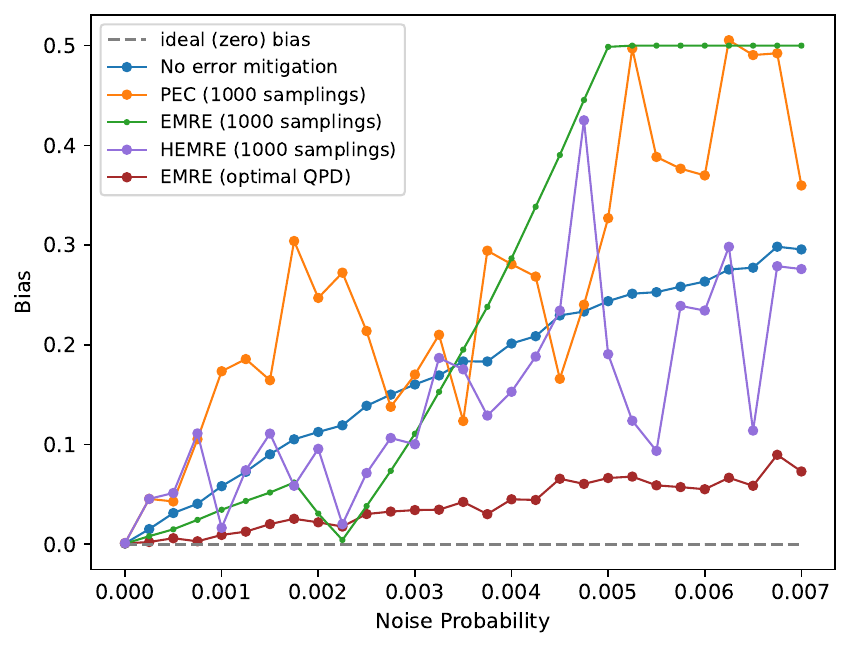}
    \caption{\textbf{Comparison of mean bias from different error mitigation protocols.} Comparison of the performance of EMRE and HEMRE against PEC and no EM under the partially depolarizing noise by plotting the mean bias in each case from the expectation values obtained in~\figref{fig:Comparison_of_expval_under_depol}. 
    We note that EMRE, performed from restricting the quasi-probabilistic decomposition obtained from PEC, outperforms PEC and HEMRE up to a certain noise probability, beyond which HEMRE continues to consistently perform better than PEC. Lastly, mitigating errors using EMRE performed using optimal generalized QPD resulted in the least bias/the best performance.
    }
\label{fig:Comparison_of_bias_under_depol}
\end{figure*}

\begin{figure}
\begin{tabular}{cc}
  \includegraphics[width=0.5\linewidth]{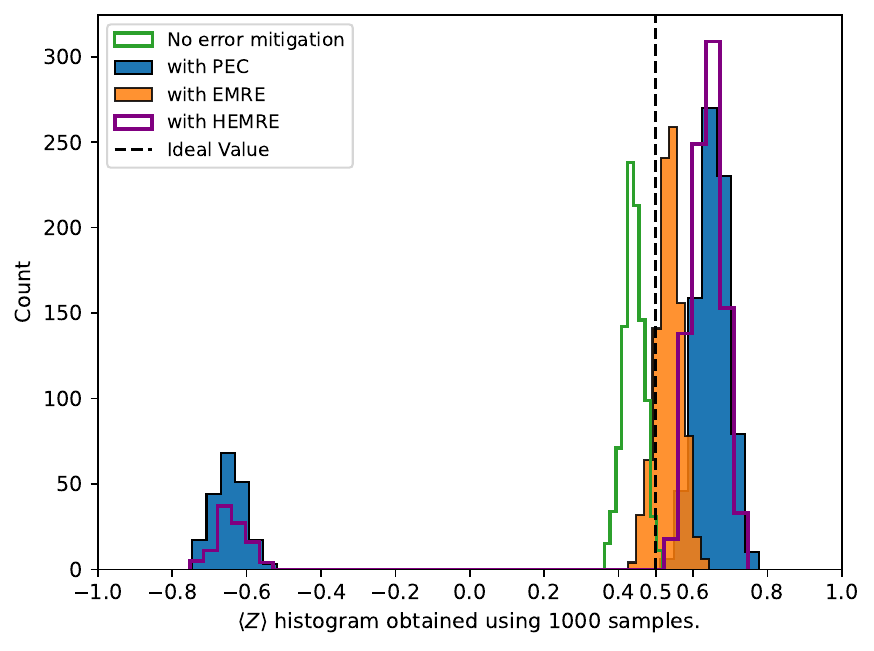} &   \includegraphics[width=0.5\linewidth]{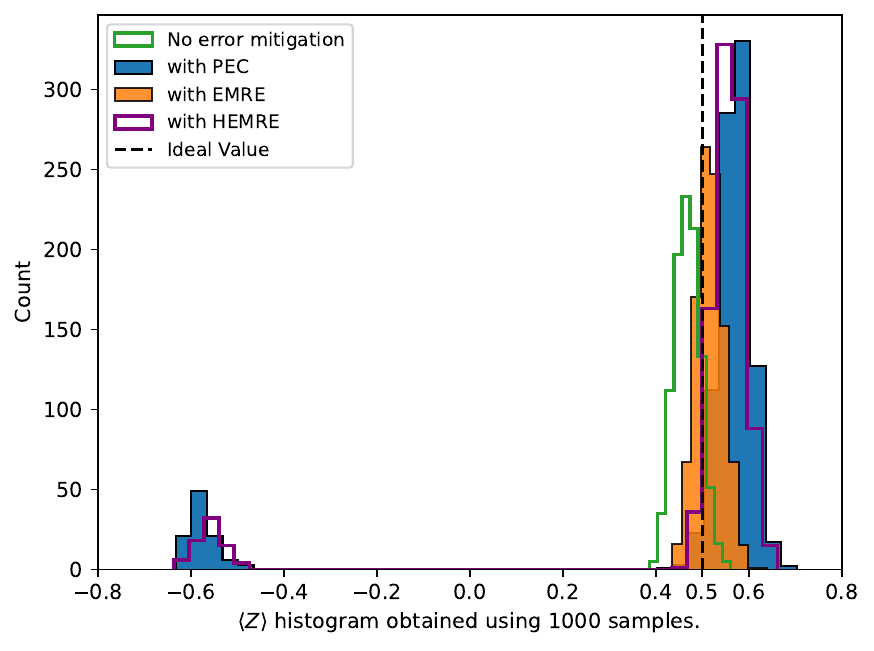} \\
(a)   & (b) \\
\end{tabular}
\caption{\textbf{Distributions of error‑mitigated expectation values.} Plots of the distribution of the 1,000 estimates of the expectation value of $Z$ obtained in the controlled-SWAP circuit without any error mitigation, with PEC, with EMRE, and with HEMRE by considering (a) depolarizing noise probability of $0.001$ and (b) depolarizing noise probability of $0.0005$. In the above plots, the tolerable bias for HEMRE was kept fixed to be 0.05.
The scattered distribution for PEC and HEMRE coming from the negative elements of the QPD is the reason behind the larger standard deviation for these two protocols. 
The ideal value is 0.5 which is marked with a black dashed vertical line.
The mean of the expectation value estimates for the noisy, PEC, EMRE and HEMRE cases for the 0.001 noise probability are 0.439, 0.392, 0.535, and 0.511, respectively, and for the 0.0005 noise probability, the respective values are 0.468, 0.455, 0.516, and 0.475. Note that there is no distribution for the optimal EMRE because it in the case of depolarizing noise, we only work with the mean of the noisy results and perform post-processing on it.
}
\label{fig:exp_val_hist_under_depol_combined_001_and_0005}
\end{figure}

The results are shown in ~\figref{fig:Comparison_of_expval_under_depol} and~\figref{fig:Comparison_of_bias_under_depol} for depolarizing noise.
We obtained 50 estimates for each of the five different cases mentioned above for each noise probability which is plotted in~\figref{fig:Comparison_of_expval_under_depol}. The shaded region indicates the standard deviation for each error mitigation. From these 50 estimates, we calculated the mean bias and plotted it against the depolarizing noise probability shown in~\figref{fig:Comparison_of_bias_under_depol}.
For estimating each expectation value, we used the same number of samplings (in this case 20 so that the total number of samplings are 1000) across different EM techniques to make a fair comparison.
PEC will give a zero bias if the number of samplings is increased.
However, the sampling overhead for PEC increases exponentially with respect to the noise probability, making it practically impossible to run any practical quantum simulation.
Therefore, PEC is often deployed with a few samples.
For EMRE, we fixed the sampling overhead and obtained `c', which only impacts the precision of estimating $E_{\mB}$ in Eq.~\ref{eq:EmB_approximation}.
So, the more samples there are, the better we estimate $E_{\mB}$. However, the bias is not impacted much as is apparent from the standard deviation in EMRE's results in~\figref{fig:Comparison_of_bias_under_depol} (or the spread of the EMRE distributions in~\figref{fig:Comparison_of_expval_under_depol} and~\figref{fig:exp_val_hist_under_depol_combined_001_and_0005}). 
For HEMRE, the number of samples needed depends on the chosen tolerable bias based on which the protocol decides which gates to approximate and which not.
In the current analyses, since we fixed the number of samples and adjusted the tolerable bias with increasing probability, we see that the mean bias that we obtained from HEMRE remained better than that of PEC for most noise probabilities. Remarkably, the standard deviation in HEMRE remained much smaller than that of PEC for large noise probabilities.

We find that the bias obtained by deploying optimal EMRE using the analytical results (brown line in \figref{fig:Comparison_of_bias_under_depol}) gives the least bias.
Specifically, the optimal bias from EMRE is about $75$ to $80\%$ better than that of PEC for all probabilities, thus outperforming PEC. Moreover, obtaining these results did not incur any extra time, as we just multiplied the mean of the noisy results with a factor followed by a few checks, to estimate the expectation value.
Note that we only have an exact result for the single-qubit gates, and for the two-qubit gates, we have used the upper bound from Eq.~\eqref{eq:ub_gen_rob_nqubit_depol}. An exact result for the two-qubit case, will lead to further reduction in the bias.
For EMRE deployed by restricting the quasi-probabilistic decomposition (green plot in~\figref{fig:Comparison_of_expval_under_depol} and~\figref{fig:Comparison_of_bias_under_depol}) and for noise probabilities below 0.003 i.e., gate fidelities to be around 99.7\% and more, EMRE outperforms HEMRE and PEC (this is where EMRE intersects PEC).
For these probabilities, we found that the bias from EMRE is about $40\%$ better than that of PEC.
Another key observation is that the error bars from EMRE (see the green EMRE plot in~\figref{fig:Comparison_of_expval_under_depol}) are very small compared to the rest, noticing that the estimates from EMRE are stable across samplings. These error bars are the spread of the 50 estimates which themselves are the mean of 20 estimates, thus the overall mean is the mean over 1,000 samples.
It also indicates that the mean of a few estimates do not spread out a lot as is the case with PEC.
In other words, EMRE demonstrates superior sample efficiency compared to PEC, particularly for low noise probabilities, when both PEC and EMRE are implemented using the same QPD.
While PEC exhibits significant bias and uncertainty for reduced sample size,  EMRE consistently yields accurate estimates of the ideal expectation value with the same sampling overhead.
For larger noise probabilities, where the bias from EMRE (green line) starts to diverge due to the growing generalized robustness of the gates, that is where HEMRE plays a part. We observe that HEMRE gave a lower bias on average than PEC for most noise probabilities while utilizing the same number of samples.
This was expected since HEMRE acquires properties of both EMRE and PEC, and approximates fewer gates based on the maximum tolerable bias provided.

EMRE and HEMREs, like PEC, rely on the generalized quasi-probabilistic decomposition of the gates for execution.
Thus, it is imperative to compare the distributions of the expectation values obtained from PEC, EMRE and HEMRE given such a decomposition and how these mitigation protocols perform given a fixed sampling overhead.
For this demonstration, let us consider the same set-up as before: the SWAP test under the local partially depolarizing noise, and we use the quasi-probabilistic decomposition for all the gates in the circuit. Note that a quasi-probabilistic decomposition is also a generalized quasi-probabilistic decomposition but it might not be the optimal one.
In Fig.~\ref{fig:exp_val_hist_under_depol_combined_001_and_0005}(a) and Fig.~\ref{fig:exp_val_hist_under_depol_combined_001_and_0005}(b), we provide the distribution of the 1000 estimates of the expectation value $\langle Z\rangle$ for the case with no EM, PEC, EMRE, and HEMRE with 0.05 tolerable bias, for noise probabilities 0.001 and 0.0005, respectively.
In this figure, each estimate for PEC, EMRE, or HEMRE is obtained after sampling once from the known quasi-probabilistic decomposition of all gates in the circuit. Thus, the mean of 1,000 estimates is the mean of 1,000 samplings.
For PEC, we used the full quasi-probabilistic decomposition for all gates, while for EMRE, we restricted the quasi-probabilistic decompositions to their positive parts, and for HEMRE, depending on the tolerable bias of $0.05$, only some gates were restricted to the positive part of their quasi-probabilistic decomposition.
We see that in both Fig.~\ref{fig:exp_val_hist_under_depol_combined_001_and_0005}(a) and~\ref{fig:exp_val_hist_under_depol_combined_001_and_0005}(b), the spread of the estimates from the noisy and EMRE case is almost the same as expected. 
When PEC or HEMRE are deployed, we see some distribution on the negative side as well due to sampling from the quasi-probability distribution, which causes the standard deviation to increase in both these cases. 
(In \textit{Supp. Info. Sec. 6} (see Fig.~S5), we have provided another similar comparison in which we plot the distribution of estimates where each estimated value for the error-mitigated case is taken to be the mean of a few samples. In such a case, we see that the spread of EMRE is narrower as compared to the noisy case, as expected, indicating reliability of the estimate.)
Furthermore, in the current comparison, since we only used 1000 samples, we did not get a near-unbiased estimate for PEC, and to get a better estimate with PEC, the number of samplings need to be increased.
However, unlike PEC, increasing the sampling size in EMRE will not decrease the bias in the result. The key difference between EMRE and PEC is that when we have a constraint on the sampling size, EMRE's results have more reliability and will have smaller biases on average as compared to PEC.
Lastly, we see that for HEMRE, the distribution is shifted towards the ideal value as some gates were approximated by HEMRE; although the variance is almost similar to that of PEC's distribution as the tolerable bias was kept small to be 0.05 and the sampling overhead was small to achieve that bias.

In \textit{Supp. Info. Sec. 6}, we also present the histograms for the measured expectation values and the unbiased estimates obtained when using PEC, EMRE, and HEMRE, under depolarizing noise. Taking the average of the unbiased estimates gives us the estimate of the expectation value.
We see that for each circuit run, both PEC and EMRE span almost the same set of expectation values.
The main difference arises in the unbiased expectation values where we need to multiply the measured expectation values with the norm and the respective sign as obtained from the decomposition.
Additionally, we'd like to remark that for the numerical demonstrations here, we have used the quasi-probabilistic decomposition used in PEC and restricted it to its positive part to perform EMRE.
Executing EMRE by restricting the quasi-probability distribution however, is not the most optimal way as shown previously, and using the optimal EMRE decomposition gives us much better bias as we can see from~\figref{fig:Comparison_of_bias_under_depol}.
In \textit{Supp. Info. Sec. 6}, we also present results obtained for the 21-qubit circuit with 462 gates, 31-qubit circuit with 692 gates, and 51-qubit circuit with 1152 gates under partially depolarizing noise, and results obtained under the inhomogeneous Pauli noise for the 7-qubit circuit with 140 gates used here.
In \textit{Supp. Info. Sec. 7}, we present results obtained in a VQE-setting. Specifically, we compare EMRE's performance with PEC in obtaining the energy landscape in the setup of computing hydrogen molecule's and transmon's ground state energy. We use the hardware-agnostic efficient SU2 ansatz for both cases. We obtain results for varying depolarizing noise probabilities.
The results for the above cases support our claims and have similar output.
We see that in all cases, either larger circuits or different noise model, EMRE performed using the optimal generalized QPD gives a much smaller bias compared to PEC.
\\

Thus, with our numerical demonstrations on circuits with different number of gates, we demonstrate the scalability of EMRE in the regime of NISQ-era and early-FTQC era computations.
We also show that when EMRE is not performed with the optimal generalized QPD, we can rely on HEMRE to get better results using lesser sampling overhead than PEC.
\\

\noindent \textbf{\large Discussion}\\
We introduce two new error mitigation protocols, namely error mitigation by restricted evolution (EMRE) and hybrid EMRE (HEMRE).
EMRE offers a constant sampling overhead at the cost of a biased estimate of the expectation value.
HEMRE, on the other hand, takes the maximum allowed bias from the user, and then selectively approximates some gates in the circuit, and outputs the estimate of the expectation value with runtime smaller than that of PEC, and bias less than or equal to the maximum tolerable bias.
We establish that the bias in EMRE is dependent on a measure called generalized robustness, and derive bounds on it under different practically significant noise models, thus finding bounds on the EMRE's biases (refer to the methods subsection for these results).
We numerically demonstrate that for the low noise probabilities that we have from current quantum hardware, EMRE produces a better estimate of the expectation value compared to PEC with limited samples.
This is particularly relevant as we are gradually advancing to the era of early fault-tolerant quantum computing, where one would combine some error correction with error mitigation techniques to perform more reliable quantum computation \cite{Huggins2021Nov,Gonzales2023Aug}. 
As the field is progressing, various companies and research groups working on quantum hardware are claiming to reduce the logical error rate within the next five years to as low as $10^{-6}$ by 2025-26 \cite{BibEntry2021May}.
In such cases, EMRE is more reliable than PEC as it requires much less sampling overhead and gives a better bias.

The efficacy of combining gate set tomography, as employed in \cite{Endo2018Jul}, with our proposed techniques to mitigate localized and non-Markovian errors in quantum computers remains an open question. Additionally, the impact of this integration on algorithmic runtime requires further investigation. A promising avenue for future research is the dynamic application of EMRE or HEMRE. By continuously approximating gates based on observed errors, we seek to predict and mitigate subsequent errors more effectively. Reinforcement learning or deep learning approaches may prove instrumental in achieving this goal.
Another direction is to eliminate the requirement of the knowledge of hardware noise, and hence save the pre-computational cost of finding the optimal GQPD. One possible way of doing this is by using the learning-based method as shown in~\cite{PRXQuantum.2.040330}.
In this work, the authors showed that while implementing PEC, one can avoid the costly pre-computation of the optimal QPD by casting it as a learning process where the optimal QPD minimizes a given loss function.
To reduce the sampling overhead, the authors propose a method called the \textit{significant-error approach} where they truncate the set of optimization parameters (quasi-probabilities) by assuming a Pauli error model consisting only of significant Pauli strings which scale polynomially with the circuit size.
This carries some similarity with EMRE in the sense that both this method and EMRE perform some sort of restriction. 
It is an open problem to explore how to incorporate similar learning-based technique to find the optimal GQPD (or the positive part) and implement EMRE, and how well the mitigation scheme will perform when such a learning process is used to get rid of the pre-computation, and restrict the inverse channel to its positive part.
Lastly, the synergy between EMRE or HEMRE and error correction holds significant potential to advance the early fault-tolerant quantum computing era.
By exploring novel error mitigation protocols that circumvent the exponential scaling of existing methods, we believe our proposals aim to accelerate progress in the field.\\

\noindent\textbf{\large Methods}\\
\noindent\textbf{Generalized Robustness}\\
Similar to the robustness measure introduced in Ref.~\cite{Takagi2021Aug} to quantify the optimal resource cost for probabilistic error cancellation with respect to a continuous set of implementable operations, in this subsection, we introduce generalized robustness inspired from the generalized robustness measure defined for quantum operations in the context of dynamical resource theories~\cite{Liu2020Feb,Liu2019Apr,Saxena2020Jun}.
Using this measure, we provide a general methodology to get the optimal generalized quasi-probabilistic decomposition of a unitary gate, and also quantify the bias in EMRE.
To elucidate the generalized robustness, we first give an overview of the quasi-probability distribution and the robustness below.

In the quasi-probabilistic decomposition, the idea is to decompose a quantum gate as a linear combination of the implementable gates with real coefficients such that the coefficients sum to one.
By keeping together all the gates with positive coefficients and all the gates with negative coefficients, the decomposition of a unitary operation $\mU$ can be expressed as follows.
\eqs{\label{eq:quasi_prob_decomposition_of_U}
\mU = q_+ \mB_+ - q_- \mB_-,}
where $q_+,\, q_- \geq 0$, $q_+ - q_- = 1$, and $\mB_+$ and $\mB_-$ represent quantum operations which are some convex combinations of the implementable gates.
Note that such a decomposition is not unique.
The PEC uses the quasi-probabilistic decomposition of all gates in the circuit to estimate the expectation value of an observable.
Owing to the negative coefficients in the quasi-probabilistic decomposition, estimating the unbiased expectation value using the Monte Carlo sampling technique requires an exponential sample size. The optimal sample complexity is achieved when the quasi-probabilistic decomposition is optimal for all gates.
The optimal decomposition is the decomposition with the minimum absolute sum of the coefficients in the decomposition.
This minimum absolute sum of coefficients is also called the optimal overhead constant and denoted by $\gamma$ in the literature~\cite{Temme2017Nov,  Takagi2021Aug,vandenBerg2023Aug}.
Denoting the set of implementable operations in $d$-dimensions by $\mI_{\mE}(d)$, the optimal overhead constant, $\gamma$, can be cast as the following optimization problem
\eqs{\gamma_{\rm opt}(\mU) = \min\left\{\sum_i |b_i| \,\middle| \, \mU = b_i \mB_i,\, b_i\in \mathbb{R},\, \mB_i\in \mI_{\mE}(d) \right\}\,.}
However, if the set $\mI_{\mE}(d)$ is continuous, getting the optimal decomposition is a hard problem.
To get the optimal decomposition, Ref.~\cite{Takagi2021Aug} outlines a strategy where one must first compute the robustness $R_{\mI_{\mE}}$, inspired by the robustness measure in quantum resource theories, by solving an optimization problem, which can then be used to get the optimal quasi-probabilistic decomposition.
Using $\gamma_{\rm opt}$, we can define another quantity called the robustness, $R_{\mI_{\mE}}$ inspired by the robustness measure in quantum resource theories.
Robustness is related to the optimal overhead constant as $\gamma_{\rm opt}(\mU) = 2R_{\mI_{\mE}}(\mU)+1$~\cite{Takagi2021Aug}.
Using robustness, we can put a bound on the optimal overhead cost of running PEC under a given noise model.

Given the above idea of quasi-probabilistic decomposition, we now elaborate on the general methodology of getting the optimal generalized quasi-probabilistic decomposition using the generalized robustness.
Let a decomposition of a unitary operation $\mU$ be as in~\eqref{eq:quasi_prob_decomposition_of_U}.
While implementing EMRE, the strategy is to approximate the ideal $\mU$ in the circuit as
\eqs{\label{eq:basic_approximation}
\mU\approx q_+ \mB_+\,.}
If we use this approximation to find the expectation value of an observable $\mO$, this translates to
\eqs{
\tr[\mO\mU(|0\rangle\langle 0|)] \approx q_+ \tr[\mO\mB_+(|0\rangle\langle 0|)],}
with an (unavoidable) error of $q_-\tr[\mO\mB_-(|0\rangle\langle 0|)]$.
However, since we are approximating the ideal gate $\mU$ with the positive part to evaluate the estimate of the expectation value and no longer using the negative component, it is not needed for the omitted term to be a convex combination of the implementable operations.
Therefore, we can decompose $\mU$ as follows:
\eqs{
\mU = s \mB - (s-1) \mN}
with $$\mB = \sum_i p_i \mB_i,$$
where $\mB_i\in \mI_{\mE}(d)$ and $\mN\in \cptp(d)$.
The set $\cptp(d)$ denotes the set of all quantum channels or completely positive and trace preserving (CPTP) maps with input and output systems to be $d$-dimensional.
By using the above decomposition, we can reduce the unavoidable error, thus getting a better estimate of the ideal expectation value.

To find the implementable operation $\mB$ closest to $\mU$, we need to find a quantum channel $\mN$ that minimizes $s-1$.
The optimal $s-1$ is often referred to as generalized robustness or global robustness in several resource theories~\cite{PRXQuantum.2.010345, PhysRevA.106.042422, PhysRevLett.115.070503, RevModPhys.91.025001, PhysRevX.9.031053, Liu2019Apr} and is used as a resource measure to quantify the value of a quantum state or channel.
In this paper, we will denote it as $R^+_{\mI_{_{\mE}}}(\mU)$ and can be cast as the following optimization problem.
\eqs{\label{eq:primal_optimization_of_gen_robustness}
\begin{aligned}
R^+_{\mI_{_{\mE}}}(\mU) = \min&\; s-1\\
 \textrm{s.t.}&\; \frac{\mU + (s-1)\mN}{s}\in \mI_{\mE}(d),\\
 &\;  s-1\geq 0,\, \mN\in \cptp(d).
\end{aligned}
}    
The dual of the above primal problem is given by 
\eqs{\label{eq:dual_optimization_of_gen_robustness}
\begin{aligned}
R^+_{\mI_{_{\mE}}}(\mU) = \sup&\; \tr[J^{\mU} \beta]-1\\
\textrm{s.t.}&\; 0\leq \tr[J^{Y}\beta]\leq 1,\\
&\;\beta\geq 0,\, Y\in \mI_{\mE}(d),
\end{aligned}
}
where $J^{\mU} (J^{Y})$ denote the Choi matrix of $\mU (Y)$ defined as $J^{\mU} := \id\otimes \mU(\Phi^+_d)$, and $\Phi^+_d:= \sum_{i,j}|i\rangle\langle j|\otimes |i\rangle\langle j|$ is the $d$-dimensional unnormalized maximally entangled state.
As a remark, note that if the set $\mathcal{I}_{\mE}(d)$ is convex, the above optimization problem becomes a semi-definite program (SDP), for which efficient classical solvers are available. 
These solvers are considered efficient as their runtime and memory requirements scale polynomially with respect to the variable dimension and the number of constraints.
However, for the above optimization problem, the variable dimension is $d=2^n$ where $n$ is the number of qubits on which $\mU$ acts.
Consequently, computing the optimal generalized robustness still incurs exponential complexity. 
Finding the optimal solution becomes even more challenging when the set $\mI_{\mE}(d)$ is not convex.

Furthermore, similar to the optimal overhead constant $\gamma_{\rm opt}$ defined as the absolute sum of the coefficients of the quasi-probabilistic decomposition, we define another quantity called the optimal generalized overhead constant $\gamma^+_{\rm opt}(\mU)$ as
\begin{equation}
\begin{split}
\gamma^+_{opt}(\mU) = \min \Bigl\{ \sum_i q_i + \sum_j n_j\,  \Big| \,  &\mU = \sum_i q_i\mB_i - \sum_j n_j \mN_j \,,\\
&\,q_i, n_j\geq 0,\, \mB_i\in \mI_{\mE}(d)\,,\\
&\,\mN_j\in \cptp(d)\Bigr\}\,.
\end{split}
\end{equation}
The optimal generalized overhead constant is related to the generalized robustness as follows 
$\gamma^+_{\rm opt}(\mU)= 2R^+_{\mI_{_{\mE}}}(\mU)+1$,
which is similar to the relation between optimal overhead constant and robustness~\cite{Takagi2021Aug}.
Thus, by first computing the generalized robustness, we can get the optimal generalized quasi-probabilistic decomposition of $\mU$ to be used for EMRE.\\

\noindent \textbf{Bounds on generalized robustness}\\
In this subsection, we derive bounds on the generalized robustness by considering specific noise models.
These bounds are crucial since the bias of EMRE depends on the generalized robustness (see Eq.~\eqref{eq:EMRE_bias_non-trivial_case}) and therefore, by deriving bounds on generalized robustness, we derive bounds on the bias in EMRE for a given noise model.
This aids in analytically determining the conditions under which EMRE is effective.

We first derive a general bound on the generalized robustness given a certain decomposition of the unitary operation, while assuming an arbitrary noise acting in the circuit.
Then, we derive bounds under the partially depolarizing noise, the single-qubit probabilistic noise, and the single-qubit partially dephasing noise.
Under $d$-dimensional depolarizing noise and qubit dephasing noise, we derive an exact expression for the generalized robustness of a unitary operation.
Practically, these equalities imply that to mitigate the above noise, we need not replace the ideal gate, rather, we can let the error act after the ideal gate instance and perform a post-processing in the form of multiplying the result by a factor followed by a few checks as given in~\figref{fig:emre_flowchart_for_expvals}, to estimate the expectation value.
Thus, these analytical results become crucial as they save time by eliminating the need to find the generalized quasi-probabilistic decomposition.

\begin{theorem}\label{thm:bound_known_decomposition}
    Given a particular decomposition $\mU= s\mB-(s-1)\mN$ where $\mB$ is a probabilistic combination of the implementable operations and $\mN$ is some quantum channel, the optimal generalized robustness is bounded by
    \eqs{
    s-1\geq R^+(\mU)\geq \frac{1}{d^2 s}\tr[\Phi^+_d J^{\mE'}]
    }
    where $\mE' = s\mB\circ \mU^{\dagger}$.
\end{theorem}
The proof is provided in \textit{Supp. Info. Section 2}.
Using the above theorem, we can get bounds on the generalized robustness of any unitary operation if a generalized quasi-probabilistic decomposition is known.

\begin{theorem}\label{thm:gen_robustness_under_depolarizing_noise}
Consider an $n$-qubit unitary $\mU$ with local partially depolarizing noise  $\mE^{\rm depol}$ of the form
    \eqs{\mE^{\rm depol}(\rho) = \left(1 - \frac{3p}{4} \right)\rho + \frac{p}{4}\left(X\rho X+ Y\rho Y + Z\rho Z\right)\,,}
    acting on each qubit after the application of $\mU$.
The generalized robustness of $\mU$ is bounded by
    \eqs{\label{eq:ub_gen_rob_nqubit_depol}
    R^+_{\rm depol}(\mU) \leq \left(\frac{4}{4-3p} \right)^n -1\,.}

    Furthermore, when we consider the unitary to be $d$-dimensional (Kraus operators can be found in~\cite{Imany2019Jul, Gokhale2019Jun}) and the noise to be $d$-dimensional depolarizing noise, then we get the following exact result for the generalized robustness of $\mU$:
    \eqs{R^+_{\rm depol}(\mU) = \frac{d^2 - 1}{d^2 + p - d^2 p}p.\label{Eq:depolarizing_bound}}    
\end{theorem}

The proof of the above theorem is given in \textit{Supp. Info. Section 3}. 
The equality in the above result, Eq.~\eqref{Eq:depolarizing_bound}, implies that if the noise in the circuit is the local partially depolarizing noise, then we can use the noisy circuit as it is and no computational time is invested in finding the optimal generalized QPD. 
Mitigating the noise in such a case just involves a simple post-processing of two steps.
First, we can obtain $E_{\mB}$ (see Eq.~\eqref{eq:EmB_approximation}) by applying the original (noisy) circuit and multiplying the result by the factor $(d^2/(d^2+ p - d^2p))^m$ where $m$ is the number of single-qubit/qudit gates in the circuit.
Next, we use the algorithm in~\figref{fig:emre_flowchart_for_expvals} which is just a series of checks to output the final estimate of the expectation value.
(The above result can also be used to upper bound the generalized robustness for the case of multi-qudit gates.)
We use the results to mitigate errors in the SWAP test (with various number of qubits and circuit sizes) under partially depolarizing noise (see Fig.~\ref{fig:Comparison_of_bias_under_depol} and \textit{Supp. Info. Sec. 6}).
As expected, we see that performing EMRE using the optimal generalized QPD outperforms EMRE done using any other decomposition. This is discussed in detail in the numerical analyses subsection.

\begin{lemma}\label{lem:gen_rob_under_probabilistic_noise}
Given a single-qubit probabilistic error channel $\mE$ of the form
    \eqs{\label{eq:prob_error_defn}
    \mE(\rho) = p\mN(\rho) + (1-p)\rho\,,}
acting on the qubit after the application of a single-qubit operation $\mU$, the generalized robustness of $\mU$ is bounded by
    \eqs{R^+(\mU) \leq \left(\frac{p}{1-p} \right)}

\end{lemma}

\begin{theorem}\label{thm:gen_rob_under_dephasing}
Given the single-qubit dephasing error $\mE^{\rm deph}$ of the form
\eqs{\label{eq:partially_dephasing_error_defn}
    \mE^{\rm deph}(\rho) = \left(1 - \frac{p}{2} \right)\rho + \frac{p}{2}\left(Z\rho Z\right)\,,}
acting on the qubit after the application of a single-qubit operation $\mU$, the generalized robustness of $\mU$ is equal to
\eqs{
R^+_{\rm deph}(\mU) = \frac{p}{2-p} \,.}  
\end{theorem}
The proofs of Lemma~\ref{lem:gen_rob_under_probabilistic_noise} and Theorem~\ref{thm:gen_rob_under_dephasing} are provided in \textit{Supp. Info. Section 4}. 
It is interesting to note that while it is challenging to find the quasi-probabilistic decomposition to perform PEC for a given error, the problem becomes slightly less challenging (and time-saving) if we are applying EMRE when the errors can be expressed as local and probabilistic quantum operations.
The reason for this is that we can get the decomposition for EMRE for the probabilistic noise from the very definition of the noise, thus saving us any extra computational time in finding the decomposition.
In such a case of probabilistic noise, we also theoretically know the upper bound on the generalized robustness using which we can estimate the sampling size for EMRE. 
Further, if the noise is a single-qubit dephasing noise, we can let the desired unitary channel be in the circuit as it is, similar to the case of d-dimensional depolarizing noise. To mitigate noise in such a specific case, we only need to multiply the estimate of the expectation value by a factor that depends on the probability of noise, and then use the algorithm in~\figref{fig:emre_flowchart_for_expvals}.
Thus, when the optimal generalized QPD and the optimal generalized robustness are exactly known analytically for some noise model, like in the above case of the partial depolarizing and the partial dephasing noise, it will further reduce the computation overhead, and we might just be able to use the noisy circuit followed by simple post-processing to estimate the ideal expectation value.

\noindent \textbf{\large Data availability}\\
All relevant data were available in the main text and Supporting Information and can be obtained from the authors upon request. Source data are provided with this paper.\\

\noindent \textbf{\large Code availability}\\
All the software used in this work is open source. No specific software was developed for this work.\\

\noindent \textbf{\large Acknowledgements}\\
We would like to thank Zhenyu Cai, Suguru Endo, Abhinav Kandala, Ying Li and Zlatko Minev, for helpful discussions. 
We thank our management executives- Kevin Ferreira, Yipeng Ji, Paria Nejat of LG Electronics Toronto AI Lab for their constant support throughout this work.
Last but not least, we are grateful to Euwern Teh of LG Electronics Toronto AI Lab for showing us how to draw beautiful quantum circuits.
Throughout our numerical computations, we used the open-source software Mitiq~\cite{mitiq} to deploy PEC in circuits.
No funding was received for this research.\\

\renewcommand{\bibsection}{\noindent \textbf{\large References}}
\bibliography{ref}

\newpage
\noindent \textbf{\large Author contributions}\\
G.S., and T.H.K. designed and performed research. G.S. did the theoretical analysis and numerical experiments. G.S., and T.H.K. analysed the data and wrote the paper.\\

\noindent \textbf{\large Competing interests}\\
The authors declare no competing financial or non-financial interests.\\

\noindent \textbf{Correspondence} and requests for materials should be addressed to Gaurav Saxena.\\ 

\onecolumngrid

\begin{table}[h]
\centering
\begin{tabular}{|c|c|c|c|}
    \hline
    Gate & Frequency & Gen. robustness & Total gen. robustness\\
    \hline
    $\mG_1$ & $n_1$ & $s_1$ & $(s_1)^{n_1}$ \\
    $\mG_2$ & $n_2$ & $s_2$ & $(s_2)^{n_2}$  \\
    $\mG_3$ & $n_3$ & $s_3$ & $(s_3)^{n_3}$ \\
    $\mG_4$ & $n_4$ & $s_4$ & $(s_4)^{n_4}$\\
    \hline
   \end{tabular}
   \caption{Information required for selecting gates to be approximated for HEMRE }\label{table:hemre_gates_gen_robustness}
\end{table}

\end{document}